\documentclass[twocolumn,English,aps,pra,10pt,superscriptaddress,floatfix]{revtex4-2}

\usepackage{times}
\usepackage{graphicx}
\usepackage{amssymb}
\usepackage{bm}
\usepackage{amssymb,amsfonts,amsmath,amsbsy,bm,t1enc,latexsym}
\usepackage{float}
\usepackage[colorlinks=true,citecolor=blue,linkcolor=magenta]{hyperref}
\usepackage[markup=blue, authormarkupposition=left]{changes}
\usepackage{url}
\usepackage[mode=text]{siunitx}
\DeclareSIUnit \dbc {dBc}
\DeclareSIUnit \dbm {dBm}
\DeclareSIQualifier\peak{p}
\usepackage{soul}
\usepackage{changes}
\usepackage{cancel}
\usepackage{times}
\usepackage{chemformula,xspace} 
\usepackage{braket}
\usepackage[capitalise]{cleveref} 

\begin{document}


\title{Nonlinear optical bistability in microring resonators for enhanced phase sensing}

\author{Patrick Tritschler}
\email{patrick.tritschler@bosch-sensortec.com}
\affiliation{Bosch Sensortec GmbH, Gerhard-Kindler Stra{\ss}e 9, Reutlingen, 72770, Germany}
\affiliation{Institute for Micro Integration (IFM), University of Stuttgart, Allmandring 9b, Stuttgart, 70569, Germany}

\author{Christian Schweikert}%
\affiliation{Institute of Electrical and Optical Communications, Pfaffenwaldring 47, 70569 Stuttgart, Germany}

\author{Rouven H. Klenk}%
\affiliation{Institute of Electrical and Optical Communications, Pfaffenwaldring 47, 70569 Stuttgart, Germany}

\author{Simon Abdani}%
\affiliation{Institute of Electrical and Optical Communications, Pfaffenwaldring 47, 70569 Stuttgart, Germany}

\author{Onur S\"ozen}%
\affiliation{Institute of Electrical and Optical Communications, Pfaffenwaldring 47, 70569 Stuttgart, Germany}

\author{Wolfgang Vogel}%
\affiliation{Institute of Electrical and Optical Communications, Pfaffenwaldring 47, 70569 Stuttgart, Germany}

\author{Georg Rademacher}%
\affiliation{Institute of Electrical and Optical Communications, Pfaffenwaldring 47, 70569 Stuttgart, Germany}

\author{Torsten Ohms}%
\affiliation{Bosch Sensortec GmbH, Gerhard-Kindler Stra{\ss}e 9, Reutlingen, 72770, Germany}

\author{Andr\'{e} Zimmermann}%
\affiliation{Institute for Micro Integration (IFM), University of Stuttgart, Allmandring 9b, Stuttgart, 70569, Germany}
\affiliation{Hahn-Schickard, Allmandring 9b, Stuttgart, 70569, Germany}

\author{Peter Degenfeld-Schonburg}%
\affiliation{Robert Bosch GmbH, Robert-Bosch-Campus 1, Renningen, 71272, Germany}

\date{\today}

\begin{abstract}

Photonic microring resonators are used in a variety of chip-integrated sensing applications where they allow one to measure transmission intensity changes upon external signals with a sensitivity that scales linearly with the Q factor. In this work, we suggest exploiting the nonlinear self-phase-modulation effect to increase the overall sensitivity by an additional gain factor appearing when the operational point of the nonlinear resonator is chosen just at the crossover from the monostable to the bistable regime. We present the theoretical idea together with a first proof of concept experiment displaying a gain factor of 22 on a chip-integrated silicon-nitride resonator.

\end{abstract}

\maketitle


\section{Introduction}
\begin{figure}[htbp]
	\centering\includegraphics[width=8.6cm]{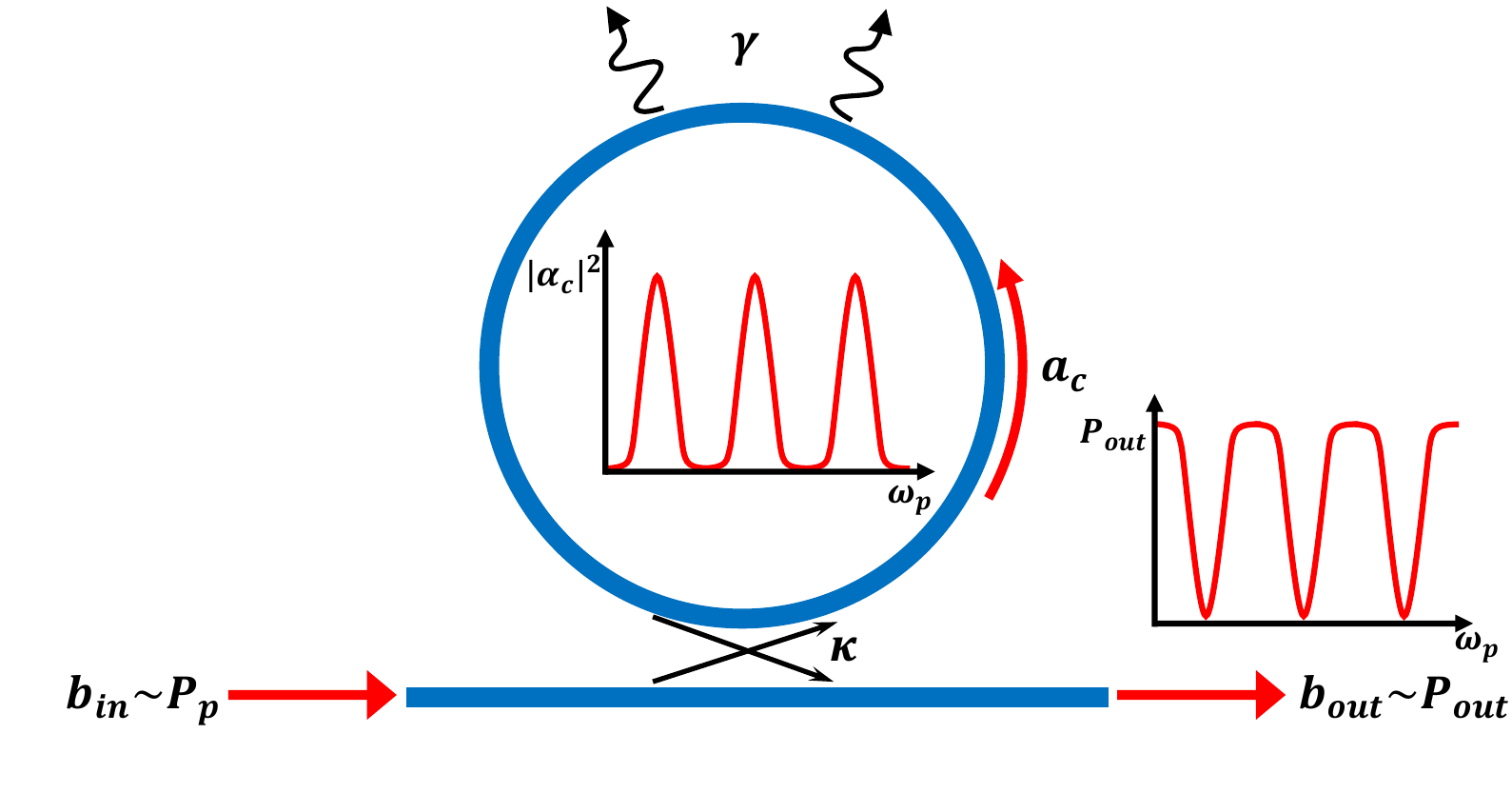}
	\caption{Schematic setup of a microring resonator. }
	\label{fig:ringresonator}
\end{figure}

Microring resonators are one of the most popular optical chip-integrated components. They have served for many fundamental investigations and pushed a vast number of technological advances \cite{Vahala2003, Armani2003, Del’Haye2007, Kippenberg2011-uh, Kippenberg2018, Chang2020, mi14051080, Kippenberg2024}. In the simplest form they consist of a ring-shaped waveguide which is within the range of the evanescent light field to a straight waveguide as illustrated in Fig. \ref{fig:ringresonator}. Ring resonators are used in many various applications like optical filters \cite{Geuzebroek2006, 588673, Zhou:07}, optical biosensors \cite{2016Biosensors, MALMIR2020164906, Toropov2021, mi14051080, Chen2020-mr}, frequency comb generation \cite{Del’Haye2007, Kippenberg2011-uh}, squeezed light generation for sensing applications \cite{tritschler2024optical, 2024_tritschler}, optical phase sensors, and many more. \\
In the following, we focus on optical phase sensors like they are used for example in ring-gyroscopes, where the Sagnac effect changes the effective path length \cite{JEOS:RP14013, photonics7040096, 10.1117/12.2304206}, or optical temperature sensors, where a temperature influence affects the ring geometry \cite{1522330, 2006Guan, ZHANG2022110494}. In both applications the optical phase and thus the resonance condition of the ring resonator are effectively changed, which also affects the output transmission power. In order to increase the sensitivity of the transmitted power with respect to a phase shift, it is necessary to increase the quality factor of the resonator by optimizing the fabrication process of the photonic devices \cite{Armani2003, Dong2009, 10.1117/12.2304206, Zhou:16}.\\
In this work, we propose to utilize the nonlinear self-phase modulation (SPM) effect to achieve a sensitivity improvement over the linear resonator operation via a multiplicative nonlinear gain factor larger than 1. The SPM effect occurs whenever the pump power and therefore the field inside the resonator are large enough to enter the nonlinear regime. Here, the nonlinear optical or thermal processes lead to the tilting of the resonance curve which even ends up in the well-known optical bistability for large enough pump power \cite{Drummond_1980}. The largest slope of the resonance curve, and thus the largest nonlinear gain factor,  appears when the operational point of the nonlinear resonator is chosen just at the crossover from the monostable to the bistable regime. We give a clear theoretical picture of how the operational point of the nonlinear resonator, also known as the Duffing or Kerr oscillator \cite{Duffing1918, Rukhlenko:10}, can be chosen in order to be at the sweet spot which can lead to nonlinear gain factors by far exceeding unity.\\
The self-heating effect in microring resonators, which leads to an SPM effect, has already been discussed in detail for chip-integrated applications, with a successful modeling in different material platforms \cite{Carmon:04, Fomin:05, 2006Johnson, Schmidt:08, 2012Arbadi, Jiang2020, ZHANG2022110494, Novarese:22, photonics10101131}. Furthermore, the SPM effect can also originate from optical four-wave mixing induced by the $\chi^{(3)}$ nonlinearity, and the combination of both effects was investigated in \cite{Ikeda:08, Chen:12}. Therefore, we experimentally characterize the thermal and optical properties of our resonator consisting of a silicon nitride (Si3N4) ring and show the increase of the sensitivity in accordance with the theoretical expectation. By placing a metal heater above the ring resonator we applied a temperature change to represent a phase shift which allowed us to measure the sensitivity improvement with an overall gain factor of up to 22. The agreement of the results to the theoretical model gives us the confidence that with a proper electronic control loop and thus with a much better control of the system parameters at the sweet spot, even gain factors of up to 100 or higher are achievable.\\
This paper is structured as follows. First, the ring resonator for the application of phase sensing is discussed in Sec. \ref{cha:microRingResonator}. The classical equations of motion to describe our system and its thermal modeling are introduced in Sec. \ref{cha:optical_modelling} and are followed by the discussion of the sensitivity boost in Sec. \ref{cha:phaseSensing}. Afterwards, the experimental setup is presented in Sec. \ref{sec:experimental-setup} followed by the results in Sec. \ref{sec:results}. We conclude with a discussion and a summary in Sec. \ref{sec:discussion} and \ref{sec:summary-and-outlook}, respectively.

\section{Nonlinear microring resonator for phase sensing}
\label{cha:microRingResonator}

In the following we introduce our model of a single nonlinear optical mode inside of a ring resonator as sketched in Fig. \ref{fig:ringresonator}. An external input laser provides the optical power $P_p$ in the straight waveguide with the amplitude $b_{\mathrm{in}}=\sqrt{P_p/\hbar\omega_p}$ and the angular frequency $\omega_p$ to pump the ring. 
At a reference temperature $T_0$, the modes inside the ring resonator have angular resonance frequencies at $\omega_{R,T_0} = 2\pi c m / n_{\mathrm{eff}}L_{\mathrm{eff}}$ with the speed of light $c$ in vacuum, the mode index $m=1,2,...$, the effective refractive index $n_{\mathrm{eff}}$, and the effective resonator length $L_{\mathrm{eff}}$ \cite{micro_ringresonators}. In the following section we first describe the optical behavior, proceed with the thermal influences, and conclude with the application for phase sensing.

\subsection{Classical dynamics of a nonlinear optical mode}
\label{cha:optical_modelling}

For our investigations it suffices to describe a single optical mode by its classical equations of motion within rotating-wave approximation \cite{Drummond_1980, tritschler2024optical}:
\begin{equation}
\begin{aligned}
\frac{d}{dt} \alpha_c &= \left( i\Delta_{tot} - \frac{\Gamma}{2} \right) \alpha_c  + \sqrt{\kappa} b_{\mathrm{in}}.
\label{equ:eqm}
\end{aligned}
\end{equation}
Here, $\alpha_c$ describes the dimensionless classical amplitude of the optical field $a_c$ inside the ring resonator in a rotating frame of the pump frequency $\omega_p$. The rate $\kappa$ describes the coupling efficiency between the waveguide and the ring, while the loss rate $\gamma$ includes all losses appearing in the ring which are summed up to the total losses $\Gamma=\kappa+\gamma$. The overall detuning between the pump frequency and the effective resonance frequency of the optical mode is given by 
\begin{equation}
\Delta_{\mathrm{tot}} = \omega_{p}-\omega_{R,T_0} + g_\mathrm{tot} |\alpha_c|^2 + \omega_{R, T_0}a_{\mathrm{th}}\Delta T_{\mathrm{ext}}
\label{equ_total_detuning}
\end{equation}
The nonlinearity of Eq.(1) appears in the detuning which depends on the photon number $|\alpha_c|^2$. This is known as the Kerr or SPM effect, and it becomes relevant when the photon number $|\alpha_c|^2$ is on the order of $\sim\omega_{R,T_0} / (g_\mathrm{tot}Q)$ with the quality factor of the resonator defined by $Q=\omega_{R,T_0} / \Gamma$. The total gain factor $g_\mathrm{tot} = g_\mathrm{opt} + g_\mathrm{th}$ includes the optical contribution $g_\mathrm{opt}$ which originates from the small but finite  $\chi^{(3)}$ nonlinearity of silicon nitride. It can be either extracted from a frequency comb measurement \cite{Kruckel:15} or directly calculated by using a quantitative value for the nonlinear refractive index \cite{tritschler2024optical}. Finally, the detuning contains contributions from thermal effects leading to both a thermal gain factor $g_\mathrm{th}$ and to a thermally induced phase shift $\omega_\mathrm{R,T_0}a_\mathrm{th}\Delta T_\mathrm{ext}$ with the externally induced temperature change $\Delta T_\mathrm{ext}$ and the the temperature coefficient $a_\mathrm{th}$. In addition to externally induced temperature changes, also self-heating causes a change of the angular resonance frequency. In this case, an optical field inside the resonator $|\alpha_c|^2$ can be absorbed by the waveguide material and leads to a heating of the system that depends on $g_\mathrm{th}$. The temperature influence is discussed in more detail in Appendix \ref{sec:thermal-modeling}.

\subsection{Sensitivity boost by the nonlinear gain factor}
\label{cha:phaseSensing}
\begin{figure}[b]
	\includegraphics[width=8.6cm]{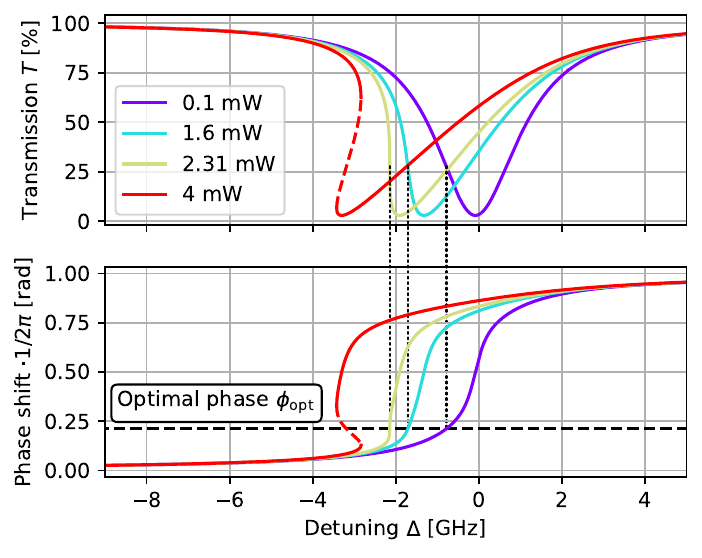}
	\caption{\label{fig:phase_dependency} 
	\textbf{Top:} Modeled output transmission for various input powers using Eq. \ref{equ:outputField} from Appendix \ref{sec:sensitivity}. The dark blue line represents the linear case at low input power, light blue shows the monostable regime below $P_\mathrm{max}$, dark yellow shows the case at $P_\mathrm{max}$, thus just at the crossover from the monostable to the bistable regime, and the red line shows the bistable regime above $P_\mathrm{max}$. \textbf{Bottom:} Phase shift of transmitted light for different input powers, highlighting the pump power independent optimal phase with the steepest transmission slope. The optimal phase in the bottom plot corresponds to the steepest slope in the top plot, marked by black dotted lines for each input power.}
\end{figure}
Next, we investigate the potential for enhancing the sensitivity of phase sensing applications by exploiting the tilting of the transmission measurement caused by the SPM effect. The conceptual idea can be understood using the steady state output transmission curves $T= P_\mathrm{out}/P_p$ over the wavelength detuning $\lambda_p -  \lambda_R$, with $\lambda_p = 2\pi c/\omega_p$ and $\lambda_R=2 \pi c/\omega_{R,T_0}$ for different pump amplitudes as shown in Fig. \ref{fig:phase_dependency}. The sensitivity of our sensing concept is directly given by the slope of the transmission curve as an external phase shift would cause a change of the resonance frequency and thus a change of the output transmission. The steady state output power $P_\mathrm{out} \equiv |b_\mathrm{out}^{ss}|^2$ is obtained from the classical input-output relation $b_\mathrm{out}^{ss} = \sqrt{\kappa} \alpha_c^{ss} - b_\mathrm{in}$, with the steady state field amplitude $\alpha_{c}^{ss} = \lim_{t\to\infty} \alpha_c(t)$. The latter is obtained by solving Eq. \ref{equ:eqm}, which however defies a simple analytical expression due to the nonlinear SPM effect as shown in more detail in Appendix \ref{sec:cavity-power}. \\
Nonetheless, we are able to put simple analytical insights for all the quantities that are required within our conceptual idea. First of all, we have the power
\begin{equation}\label{equ:maxPower}
P_\mathrm{max} =\frac{\sqrt{3}}{9}\frac{\hbar\omega_{p} \Gamma^3}{ g_{\mathrm{tot}}\kappa}
\end{equation}
which marks the crossover from the monostable to the bistable regime. Our sensing scheme will work best slightly below $P_\mathrm{max}$ and break down in the bistable regime. We find that the point of maximal slope of the transmission curve is always at a frequency detuning of $\Delta_\mathrm{opt} =  -3g_{\mathrm{tot}}\kappa P_p/\Gamma^2\hbar\omega_{p} - \Gamma/\sqrt{12} $ with $\Delta_\mathrm{opt} = \omega_p -\omega_{R,T_0}$. Finally, as shown in more detail in Appendix \ref{sec:sensitivity}, the phase sensitivity $S=\delta P_\mathrm{out} / \delta\omega$ which describes a change of the output power with respect to frequency change in W/Hz is given by
\begin{equation}\label{equ:sensitivity}
S= S_\mathrm{lin} \cdot \frac{P_\mathrm{max}}{P_\mathrm{max} - P_p} 
\end{equation} 
with $S_\mathrm{lin} = -\sqrt{27} P_p Q[1/4 - (\kappa/\Gamma-1/2)^2] / \omega_{R, T_0} $. As expected, the linear sensitivity $S_\mathrm{lin}$ scales linearly with the pump power $P_p$ and the quality factor of the resonator. Moreover, it is desired to design the ring resonator system slightly undercoupled where $\kappa<\gamma$ in order to maximize the sensitivity. Physically, a high quality factor and a design close to the critical coupling lead to a small bandwidth in combination with a large extinction ratio of the output power and thus, to a steep slope of the transmission curve, which results in a high power change at a certain frequency variation. In contrast to the linear resonator concept, the sensitivity  will be boosted by an additional multiplicative gain factor $G=P_\mathrm{max}/(P_\mathrm{max}-P_p)$ in our nonlinear resonator concept. If $P_p$ is close to $P_\mathrm{max}$, then a large nonlinear gain can be achieved. It is clearly visible in Eq. \ref{equ:maxPower} and \ref{equ:sensitivity} that a high $g_\mathrm{tot}$ results in a lower $P_\mathrm{max}$ and thus in a larger sensitivity at lower $P_p$. 
\begin{figure*}
	\centering\includegraphics[width=18cm]{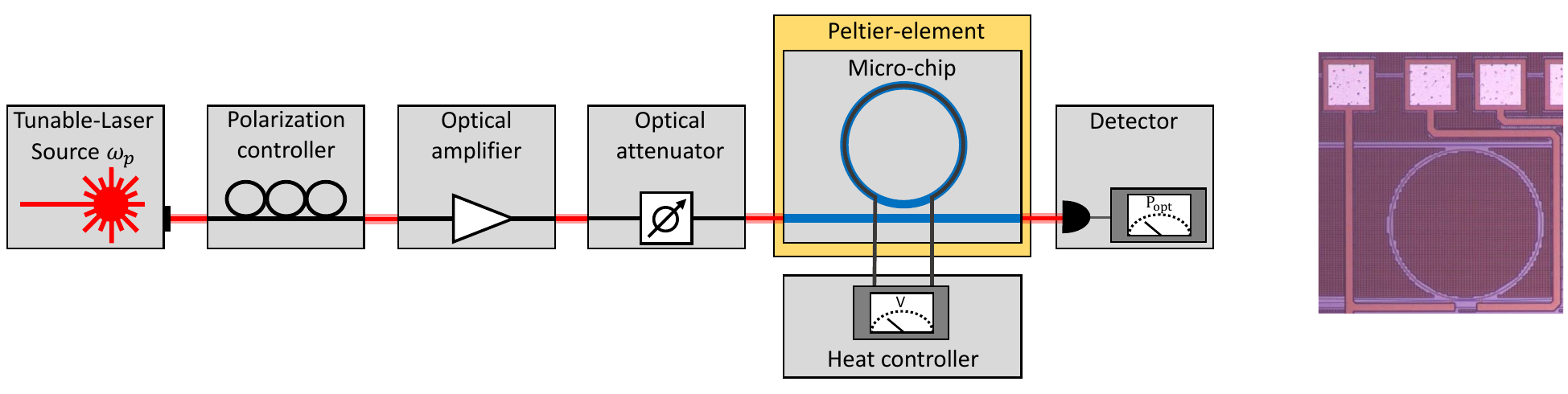}
	\caption{\label{fig:setup} \textbf{Left:} Schematic of the measurement setup. \textbf{Right:} Image of the used microring resonator structure with a metal-heater layer on top.}
\end{figure*}

\section{Experimental setup}
\label{sec:experimental-setup}
To validate the derived equations and to show the potential of our concept, we designed and characterized a chip-integrated $\mathrm{Si_3N_4}$ ring resonator. The ring consists of a waveguide with a height of 800 nm, a width of 1.6 \textmu$\mathrm{m}$, a length of 1319.46 \textmu$\mathrm{m}$ and a gap between the ring and the straight waveguide of 0.52 \textmu$\mathrm{m}$. The optical waveguide is surrounded by silicon dioxide ($\mathrm{SiO_2}$). Starting from a tuneable laser source with a center wavelength of 1550 nm, the laser light is coupled via optical fibers to a polarization controller. This is followed by an optical amplifier, an optical attenuator, and finally, the laser light is coupled into the chip by edge coupling via lensed fibers. The schematic of the experimental setup and an image of our ring resonator are shown in Fig. \ref{fig:setup}. By using the amplifier, a constant optical power is set and adjusted by the attenuator. This ensures that a polarization set by the polarization controller remains constant as the power varies. The light in the optical chip couples from the straight waveguide into the ring resonator, and the out-coupled light is collected together with the transmitted one via edge coupling and then sent to an optical power meter which forms the detector stage. The optical power is measured before and after the chip to determine the coupling losses of the system. A metal heater is placed above the ring resonator and is connected to a voltage controller to be able to change the temperature of the ring. By using this heater, the resonance frequency $\omega_{R, T_0}$ can be influenced with an applied temperature shift and represents a possible phase shift to mimic an external phase shift which we exploit to measure the sensitivity of our setup. For an additional thermal characterization, a Peltier element is placed below the microchip to be able to apply an accurate temperature change to the whole chip for further characterization.

\section{Results}
\label{sec:results}
\begin{figure}[b]
	\includegraphics[width=8.6cm]{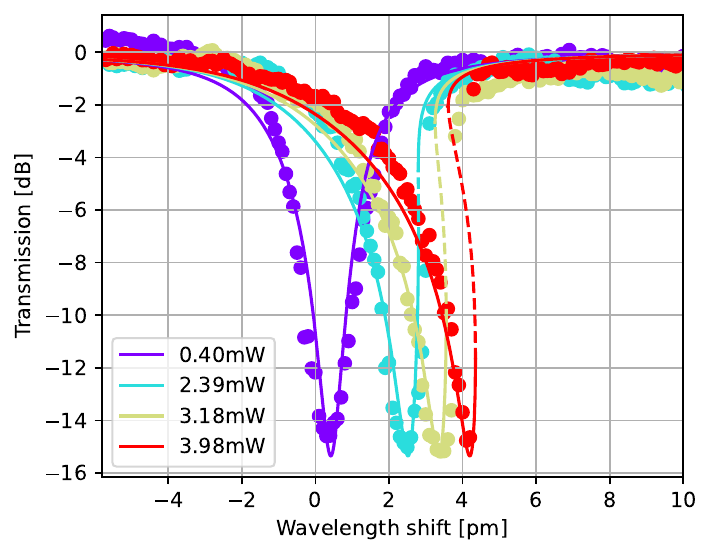}
	\caption{\label{fig:bistability_meas} Optical bistability measurements (dots) and the theoretical model using the experimentally determined parameters (lines).
	}
\end{figure}

\begin{figure}[b]
	\includegraphics[width=8.6cm]{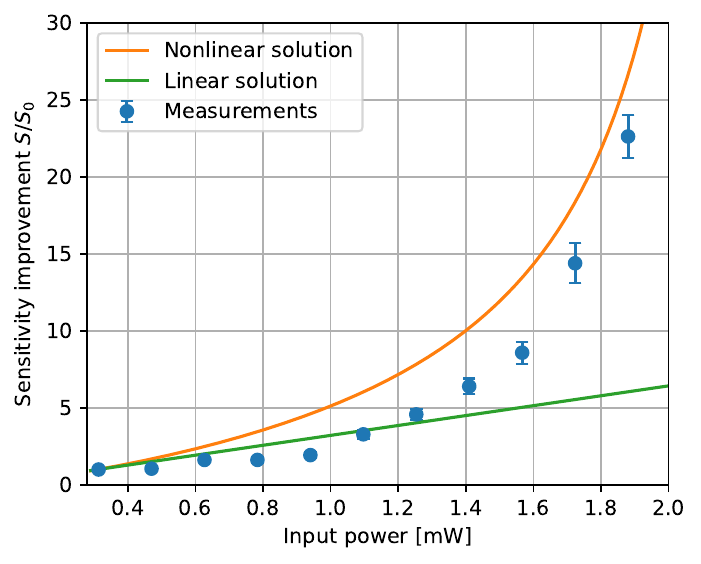}
	\caption{\label{fig:sensitivity_meas_calc} Sensitivity measurements (blue), the modeled expectation for the linear model (green), and the nonlinear model according to Eq. \ref{equ:sensitivity} (orange).}
\end{figure}

First, we vary the temperature and measure the transmitted power to characterize the thermal properties of the ring resonator. Therefore, we set a fix temperature by the Peltier element and vary it to measure the change of the resonance frequency from which we identify $a_\mathrm{th}= 1.226\cdot10^{-5}$ 1/K. \\
Next, we perform transmission measurements by increasing the wavelength and starting from a low optical power inside the straight waveguide with $P_p=0.40$ mW to characterize the linear behavior from which we deduce the coupling rate $\kappa=1450.64$ MHz as well as the loss rate to $\gamma= 1027.42$ MHz from the classical input-output theory which is shown in Eq. \ref{equ:outputField} in the Appendix. This leads to an overcoupled ring resonator with a Q factor of $4.9\cdot10^{5}$. Then, we increase the power up to where the bistability appears and use the transmission curves to extract a total gain factor of $g_\mathrm{tot} \approx 112$ Hz. The optical gain factor is calculated to $g_\mathrm{opt} \approx 2$ Hz, which leads to $g_\mathrm{th} \approx 110$ Hz, and this shows that self-heating is the dominant process over the optical SPM effect. The measurement results and the calculations using the extracted parameter values are shown as dots and lines in Fig. \ref{fig:bistability_meas}, respectively. The tilting of the transmission spectrum and a bistability at increased input power is clearly visible while the theoretical model matches well with the measurements. \\
To demonstrate the sensitivity improvement by utilizing the SPM effect, we perform sensitivity measurements for different input powers while keeping $P_p$ below $P_{\mathrm{max}}$ which using the extracted parameter values results in $P_{\mathrm{max}}=2.31$ mW. For this purpose, we set the wavelength of the pump laser to match the optimal detuning $\Delta_{\mathrm{opt}}$ for each pump power and applied additional small phase shifts using the metal heater above the ring resonator. Finally, we measured the change of the optical output power to determine the sensitivity $S$. The results are shown in Fig. \ref{fig:sensitivity_meas_calc}. Each value at various input power $P_p$ is normalized to the measurement result at a small input power with the reference sensitivity $S_0$ which corresponds to the linear resonance curve. It can be seen that the sensitivity rises with an increasing input power. Clearly, the measurement results follow the trend of the sensitivity as predicted by the theoretical model with Eq. \ref{equ:sensitivity}. Additionally, the linear trend is shown as a reference. We will discuss the deviation of the measured results from the theoretical predictions in the next section. 

\section{Discussion}
\label{sec:discussion}

The results displayed in Fig. \ref{fig:sensitivity_meas_calc} show a sensitivity improvement factor of up to 22.6 in contrast to the reference sensitivity $S_0$ in the linear operation limit of the ring resonator. This corresponds to a linear enhancement of $S_\mathrm{lin}=5.7$ and an additional nonlinear gain of $G=3.97$. Note, that we only managed to properly measure the slope of the transmission curves for input powers of up to ~1.9 mW which corresponds to only 80 \% of the maximal power $P_\mathrm{max}\approx2.3$ mW. This is due to the fact that with our setup we were only able to resolve the detuning between the pump wavelength and the effective resonance wavelength of the cavity with an accuracy of 0.1 pm. We expect that with a higher resolution an operation close to $P_\mathrm{max}$ is possible and the overall nonlinear gain factor will be much more convincing.\\
Moreover, we have found that the phase shift between the input laser light and the output light is independent of the input power and constant at the optimal frequency detuning $\Delta_\mathrm{opt}$; see also Fig. 2. At $\Delta_\mathrm{opt}$ the phase shift is given by 
\begin{equation}\label{equ:phase_shift}
\phi_{\mathrm{opt}} = -i\log\left(-\frac{2\Gamma-3\kappa + \sqrt{3} i \kappa }{\sqrt{   3\kappa^2 + \left( 2\Gamma - 3\kappa \right)^2 }}  \right) + \pi.
\end{equation}
Therefore, one can use an electronic control loop to find the laser frequency or rather wavelength which fulfills the phase shift $\phi_\mathrm{opt}$ to be just at the optimal operational point of steepest slope and thus highest sensitivity. For this operation, it is also necessary to use a photonic package to realize a power-stable operation. If these measures are all implemented, we expect that nonlinear gain factors of up to 100 or even higher are achievable. \\ 
Finally, we discuss the impact of noise on the sensor performance. There are several sources of noise with the most dominant ones arising from the laser source, the sensor area, the detector, and the electronic circuit. The noise that is already present in signals entering the resonator is also amplified as the sensitivity increases. However, in a well designed sensor system with a proper control loop, the dominant noise arises from the detector \cite{Zhou:16, Cahillane:21} and the increase in sensitivity actually reduces the detector noise as shown in detail in Appendix \ref{sec:noise}. This improves the so-called limit of detection (LOD) with the nonlinear gain. Thus, our method improves both, the sensitivity as well as the LOD for optical sensor systems where the detector noise is dominant.

\section{Summary}
\label{sec:summary-and-outlook}

In this work we demonstrate the possibility of using the tilting of the transmission measurement caused by the SPM effect to increase the sensitivity of a ring resonator for phase sensing applications. We introduced the theoretical concept on a single optical mode with a Kerr and self-heating nonlinearity and validated our model by measurements on a chip-integrated $\mathrm{Si_3N_4}$ ring resonator where we have been able to prove a sensitivity improvement factor of around 22 experimentally.\\
However, our sensing concept gives the perspective that with a well-suited control loop sensitivity improvement factors of more than 100 are achievable. While high quality factors are reachable with advanced process technology, our approach requires advanced system control to reach the same goal. In addition, combining high Q factors with our approach promises to reach far beyond state of the art sensitivities. Thus, we provide a new measure toward chip integrated sensor solutions with implications on size, cost, and performance in many possible applications such as temperature sensors, biosensors, and optical gyroscopes.

\begin{acknowledgments}
	The IPCEI ME/CT project is supported by the Federal Ministry for Economic Affairs and Climate Action on the basis of a decision by the German Parliament, by the Ministry for Economic Affairs, Labor and Tourism of Baden-W\"urttemberg based on a decision of the State Parliament of Baden-W\"urttemberg, the Free State of Saxony on the basis of the budget adopted by the Saxon State Parliament, the Bavarian State Ministry for Economic Affairs, Regional Development and Energy and financed by the European Union - NextGenerationEU.
\end{acknowledgments}

\appendix

\section{Thermal modeling}\label{sec:thermal-modeling}

In this Appendix we derive the detuning $\Delta_{\mathrm{tot}}$ for the main text in more detail which is required to model the SPM effect. If temperature is applied to the ring resonator, then the effective index as well as the geometric length of the resonator changes, which leads to a shift of the angular resonance frequency and depends on the temperature coefficient $a_\mathrm{th}= \frac{1}{L_{\mathrm{eff}}}\frac{dL_{\mathrm{eff}}}{dT} + \frac{1}{n_{\mathrm{eff}}}\frac{dn_{\mathrm{eff}}}{dT}$ \cite{Carmon:04}. However, in addition to externally induced temperature changes, also self-heating causes a change of the angular resonance frequency. In this case, an optical field inside the resonator can be absorbed by the waveguide material and leads to a heating of the system due to different inelastic photon-phonon interaction processes like surface state absorption which also appears in $\mathrm{Si_3N_4}$ ring resonators \cite{Agarwal1996, Levy2010} such as the one used in our experiment. Following \cite{Fomin:05, Haus1984}, we start with the equation of thermal diffusion to describe the dynamics of the internal temperature change $\Delta T_{\mathrm{int}}$ caused by self-heating inside the ring resonator with 
\begin{equation} 
\frac{d}{dt} \Delta T_{\mathrm{int}}(t) = \delta_{\mathrm{th}} \hbar \omega_p |\alpha_{c}|^2 - \gamma_{\mathrm{th}} \Delta T_{\mathrm{int}}(t). \label{equ:thermalDiffusion}
\end{equation}
Here, the thermal relaxation rate $\gamma_{\mathrm{th}}$ corresponds to the heat equalization with the environment and the thermal absorption rate $\delta_{\mathrm{th}}$ describes the self-heating of the ring resonator. The thermal absorption rate is given by
\begin{equation}
\delta_{\mathrm{th}} = \frac{2n_{\mathrm{eff}}\gamma_{\mathrm{abs}}}{c_p \rho A_{\mathrm{eff}}L_{\mathrm{eff}}},
\end{equation}
with the specific heat capacity $c_p$, mass density $\rho$, and the loss rate $\gamma_{\mathrm{abs}}$ that depends on the absorption losses $\alpha_{\mathrm{abs}}$ with \cite{ZHANG2022110494}
\begin{equation}
	\gamma_{\mathrm{abs}} = \frac{\left(1-e^{-\alpha_{\mathrm{abs}}L_\mathrm{eff}}\right)c}{n_\mathrm{eff}L_\mathrm{eff}}
\end{equation}
The thermal relaxation rate $\gamma_{\mathrm{th}}$ depends on the thermal capacity $C_{\mathrm{th}}=\rho c_p V$ and the thermal resistance which can be approximated for a rectangular waveguide to $R_{\mathrm{th}}=1/4kL_{\mathrm{eff}}$ with the thermal conductivity $k$ and the waveguide volume $V$ \cite{Fomin:05, Gu:14, Haus1984}. Using $A_{\mathrm{eff}}\approx V /L_{\mathrm{eff}}$ we get 
\begin{equation}
\gamma_{\mathrm{th}} = \frac{1}{\tau_{\mathrm{th}}} = \frac{1}{R_{th}C_{\mathrm{th}}} = \frac{4k}{\rho c_p A_{\mathrm{eff}}}.
\end{equation}
which in our work results in the thermal relaxation time $\tau_{\mathrm{th}}\approx0.5$ \textmu$\mathrm{s}$. It is important to note that a correct assumption for $\gamma_{\mathrm{th}}$ is not possible using only the values of the optical waveguide due to the complex interaction between the ring resonator and the materials around it. Instead, a better approximation is achieved by using the box material for $k$, as the material around the optical waveguide plays a crucial role in heat dissipation \cite{photonics10101131}. Solving Eq. \ref{equ:thermalDiffusion} for the steady state leads to 
\begin{equation}
\Delta T_{\mathrm{int}} = \frac{\delta_{\mathrm{th}}\hbar \omega_p}{\gamma_{\mathrm{th}}} |\alpha_{c}|^2 = \frac{n_{\mathrm{eff}} \gamma_{\mathrm{abs}} \hbar \omega_p}{2kL_{\mathrm{eff}}} |\alpha_{c}|^2.
\end{equation}
The thermal detuning can be introduced using this result, the definition of an external temperature change $\Delta T_{\mathrm{ext}}$ with  $\Delta T = \Delta T_\mathrm{ext} + \Delta T_\mathrm{int}$ as 
\begin{equation}
\Delta_{\mathrm{th}} = \omega_{p} - \omega_{R,T_0} + \omega_{R,T_0}a_{\mathrm{th}}\Delta T_{\mathrm{ext}} + g_{\mathrm{th}} |\alpha_{c}|^2
\end{equation}
with the the thermal gain
\begin{equation}
g_{\mathrm{th}} = \frac{\delta_{th} \hbar \omega_p a_{\mathrm{th}} \omega_{R,T_0}}{\gamma_{\mathrm{th}}} \approx \frac{\hbar \omega_p^2 n_{\mathrm{eff}} \gamma_{\mathrm{abs}} a_{\mathrm{th}}}{2kL_{\mathrm{eff}}}.
\label{equ:thermal_gain}
\end{equation}
Finally, combining the optical and the thermal detuning leads to the total detuning which includes both influences as
\begin{equation}
\Delta_{\mathrm{tot}} = \omega_{p}-\omega_{R,T_0} + g_{\mathrm{opt}}|\alpha_{c}|^2 + g_{\mathrm{th}} |\alpha_{c}|^2 + \omega_{R, T_0}a_{\mathrm{th}}\Delta T_{\mathrm{ext}}
\label{equ_total_detuning_app}
\end{equation}
The optical nonlinearity is defined as 
\begin{equation}
g_{\mathrm{opt}} \approx \frac{\hbar \omega_p^2 v_g^2 n_2}{c A_{\mathrm{\mathrm{eff}}}L_{\mathrm{eff}}},
\end{equation}
with the group velocity $v_g$, the effective mode area $A_{\mathrm{eff}}$, and the nonlinear refractive index $n_2$ \cite{PhysRevA.92.033840,	tritschler2024optical}.

\section{Cavity power}\label{sec:cavity-power}

To derive the equations in the main text, we start by describing the steady state field amplitude inside of the resonator $\alpha_{c}^{ss}$. Using Eq. \ref{equ:eqm} of the main text with $\Delta_\mathrm{tot}$ for the detuning, we can introduce three dimensionless parameters for the cavity field $A$, the pump field $B$, and the detuning $\delta$ with the definitions
\begin{eqnarray}
|A|^2 &=& \frac{g_{\mathrm{tot}}}{\Gamma} |\alpha_{c}|^2, \label{equ:a_norm} \\
|B|^2 &=& \frac{g_{\mathrm{tot}}}{\Gamma^3} \kappa |b_{in}|^2, \\
\delta &=& \frac{\Delta_{\mathrm{tot}}}{\Gamma} = \frac{\Delta_{\mathrm{lin}}}{\Gamma} + \frac{g_{\mathrm{tot}}}{\Gamma} |\alpha_{c}|^2 = \tilde{\delta} + |A|^2. \label{equ:d_norm}
\end{eqnarray}
Thereby, $g_{\mathrm{tot}}=g_{\mathrm{th}} + g_{\mathrm{opt}}$ is the total nonlinear gain and $\Delta_{\mathrm{lin}}=\omega_{p} - \omega_{R,T_0} + \omega_{R, T_0}a_{\mathrm{th}}\Delta T_\mathrm{ext}$ the linear detuning with $\tilde{\delta}=\Delta_{\mathrm{lin}} / \Gamma$. This leads to the following dimensionless description of the field inside of the ring resonator:
\begin{equation}
|A|^2 = \frac{|B|^2}{\frac{1}{4} + \left(\tilde{\delta} + |A|^2 \right)^2}.
\label{equ:dimensionlessIntracavity}
\end{equation}
The motivation for the dimensionless description is that this leads to simpler solutions that only depend on $B$ and $\tilde{\delta}$. Solving Eq. \ref{equ:dimensionlessIntracavity} for $|A|^2$ leads to one physical solution for the input power at $	|B|^2 \leq \sqrt{3}/9$ given by
\begin{equation}
|A|^2 = \frac{3 - \left( 2\tilde{\delta} + \sqrt[3]{\Xi}  \right)^2}{6\sqrt[3]{\Xi} }
\label{equ:dimensionlessIntracavity_solved}
\end{equation}
with
\begin{eqnarray}\label{equ:normed_intracavity}
\begin{aligned}
\Xi &= -8\tilde{\delta}^3 - 18\tilde{\delta} - 108 B^2 +  3\sqrt{3}&\\
&\cdot  \sqrt{16\tilde{\delta}^4 + 64\tilde{\delta}^3B^2 + 8\tilde{\delta}^2 + 144\tilde{\delta}B^2 + 432B^4 + 1}.&
\end{aligned}
\end{eqnarray}
The crossover from the monostable to the bistable regime is marked by $|B|^2 = \sqrt{3}/9$ which in physical units leads to the definition of the maximum power as introduced in Eq. \ref{equ:maxPower} of the main text. Using Eq. \ref{equ:dimensionlessIntracavity_solved}, we can then also determine the optimal detuning frequency for the steepest slope to
\begin{equation}\label{equ:delta_opt_n}
	\tilde{\delta}_{\mathrm{opt}}= - 3|B|^2 -1/\sqrt{12}.
\end{equation}
In physical units the linear detuning at the steepest slope becomes $\Delta_\mathrm{opt} \equiv \delta_{\mathrm{opt}} \Gamma$ and thus as stated in the main text with the following equation
\begin{equation}
	\Delta_\mathrm{opt} =  -\frac{3g_{\mathrm{tot}}\kappa}{\Gamma^2} \frac{ P_p}{\hbar\omega_{p}} - \frac{\Gamma}{\sqrt{12}}.
\end{equation}
Using Eq. \ref{equ:dimensionlessIntracavity_solved} and \ref{equ:delta_opt_n}, it becomes clear that at the point of the steepest slope, we always have $|A|^2 = 3|B|^2$ and therefore also $\tilde{\delta} + |A|^2 = -1/\sqrt{12}$.

\section{Sensitivity of the ring resonator}\label{sec:sensitivity}

To determine the sensitivity of the system, we first analyze the output field of the resonator. Following \cite{Collet84, Gardiner85} and using the input-output theory with $b_{\mathrm{out}} = \sqrt{\kappa} a_{c} - b_{\mathrm{in}}$, we can then solve Eq. \ref{equ:eqm} for the steady state and derive the equation of the output-field $b_{\mathrm{out}}$ in dependency of the input field to
\begin{equation} 
b_{\mathrm{out}} = \left( \frac{\kappa }{\frac{\Gamma}{2} - i \Delta_{\mathrm{tot}}} -1 \right) b_{\mathrm{in}}.
\label{equ:outputField}
\end{equation}
This equation is used to fit the measured transmission spectrum from Fig. \ref{fig:bistability_meas}. Using this equation \ref{equ:outputField} together with the dimensionless parameters from the Eqs. \ref{equ:a_norm}-\ref{equ:d_norm}, we can derive the output power to
\begin{equation}\label{equ:normed_Pout}
	P_\mathrm{out} = P_{p} \cdot \frac{\left( \frac{\kappa}{\Gamma}-\frac{1}{2} \right)^2 + \left( \tilde{\delta} + |A|^2 \right)^2 }{\frac{1}{4} + \left(\tilde{\delta} + |A|^2 \right)^2 }.
\end{equation}
In a sensor application, the resonance condition and thus the detuning $\tilde{\delta}$ shifts due to external signal influences which represent basically a phase shift. Similar to Eq. \ref{equ:d_norm}, we can introduce the phase shift $\phi=\delta\Delta/\Gamma$, and we can set $\tilde{\delta} \to \tilde{\delta} + \phi$, where $\phi$ originates from an externally induced phase shift or an externally induced temperature change with $\phi\sim\Delta T_\mathrm{ext}$. Then, the sensitivity is defined by the linearization of Eq. \ref{equ:normed_Pout} with respect to $\phi$ as $P_\mathrm{out}(\phi)=P_\mathrm{out}(0) + S\delta\Delta$ with
\begin{equation}\label{taylor_series}
S= P_{p} \cdot G_\mathrm{geo} G_\mathrm{det} \left( 1 + \frac{d}{d \phi} |A|^2 \right).
\end{equation} 
Thereby, $ G_\mathrm{geo}$ describes the geometric gain with
\begin{equation}\label{equ:geometric_gain}
	 G_\mathrm{geo} = \frac{Q}{\omega_{R,T_0}} \left[ \frac{1}{4} - \left( \frac{\kappa}{\Gamma} - \frac{1}{2} \right)^2  \right]
\end{equation}
with the $Q$ factor of the ring resonator $Q=\omega_{R,T_0}/\Gamma$. The detuning gain $G_\mathrm{det}$ in Eq. \ref{taylor_series} is given by
\begin{equation}
	G_\mathrm{det} = \frac{2\left( \tilde{\delta} + |A|^2  \right)}{\left[\left( \tilde{\delta} + |A|^2 \right)^2 + \frac{1}{4} \right]^2}
	\xrightarrow[]{\tilde{\delta}=\tilde{\delta}_\mathrm{opt}}  -\sqrt{27}.
\end{equation}
It is clearly visible, that the best $G_\mathrm{geo}$ can be achieved with a high $Q$-factor and $\kappa/\Gamma=1/2$, which is the case for the so-called critical coupling of the ring resonator. \\
To arrive at Eq. \ref{equ:sensitivity} of the main text, we finally have to show that $1+d|A|^2/d\phi$ reduces to $P_\mathrm{max}/(P_\mathrm{max}-P_p)$ at the optimal detuning $\tilde{\delta} = \tilde{\delta}_\mathrm{opt}$. Therefore, we put $\tilde{\delta} = \tilde{\delta}_\mathrm{opt} +\phi$ into Eq. (B4) and take the derivative with respect to $\phi$ to arrive at
\begin{equation}\label{equ:part}
\frac{d}{d\phi} |A|^2  = -\frac{2|B|^2 \left(\tilde{\delta}_\mathrm{opt} + \phi + |A|^2\right)}{\left[\left( \tilde{\delta}_\mathrm{opt}+\phi + |A|^2 \right)^2 + \frac{1}{4} \right]^2} \cdot  \left(1+\frac{d}{d\phi} |A|^2\right).
\end{equation}
Next, we evaluate Eq. \ref{equ:part} at $\phi=0$, exploit that $\tilde{\delta}_\mathrm{opt} + |A|^2 = -1/\sqrt{12}$ and realize the relation $P_p/P_\mathrm{max} = 9|B|^2/\sqrt{3}$ to arrive at Eq. \ref{equ:sensitivity} of the main text after a few more simple algebraic steps. 

\section{Noise analysis}\label{sec:noise}
\begin{figure}
	\includegraphics[width=8.6cm]{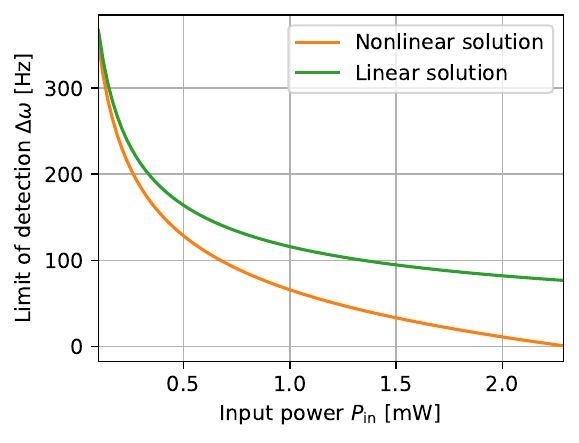}
	\caption{\label{fig:LOD} Calculated limit of detection for the linear model (green) and the nonlinear model (orange) according to Eq. \ref{equ:minDetFreqRes} using the same parameters as in the main text with $\eta_d=1$ and an integration time of 10 ms which corresponds to a detector bandwidth of $T^{-1}=100$ Hz.
	}
\end{figure}
To determine if the proposed nonlinear gain is advantageous in sensor applications, it is very important to discuss its impact on noise sources inside the system. Therefore, the LOD, which corresponds to the minimum detectable signal change, is analyzed in the following. \\
As mentioned in the main text, the main noise arises from the laser source, the sensor area, the detector, and the electronic circuit. These noise sources are combined and influence the LOD. Thereby, the noise that is already present in signals entering the ring resonator is amplified by the nonlinear gain, and thus, a sensor system that is dominated by these noise sources cannot be improved by the nonlinear gain. However, in well designed optical sensor systems with a proper control loop, the detector noise is the dominant one which can be reduced in the same manner by the quality factor \cite{Zhou:16, Cahillane:21} and also by the nonlinear gain. Especially, the control loop is required to deal with the phase and frequency noise of the laser source. We derive the LOD by following state of the art procedures \cite{BRAUNSTEIN_phaseLimit_2, DEMKOWICZDOBRZANSKI2015345}, where a photo detector measures the intensity, thus the output photon number via the integrated output photon flux operator $\hat{D}= \int_0^T dt \hat{b}_\mathrm{out}^+(t) \hat{b}_\mathrm{out}(t)$ over the integration time $T$. The lower frequency resolution $\Delta\omega$ is given by error propagation with
\begin{equation}
	\Delta\omega = \frac{\sqrt{\Delta D}}{|\delta \langle D \rangle / \delta \omega|}.\label{equ:errorPropagation}
\end{equation}
Eq. \ref{equ:errorPropagation} states that the minimum detectable frequency change $\Delta\omega$ is determined by the ratio between the variance $\Delta D = \langle D^2 \rangle - \langle D \rangle^2$ and the detector sensitivity $\delta\langle D \rangle / \delta \omega$. For a coherent state it can be easily shown that $\langle D \rangle = \Delta D$ with
\begin{equation}
	\Delta D = \frac{\eta_d T P_\mathrm{out}}{ \hbar \omega_p},
\end{equation}
with the detector efficiency $\eta_d$. Thus the sensitivity $S=\delta P_\mathrm{out} / \delta \omega$ which we have introduced above and in the main text enters the LOD via $\delta \langle D\rangle / \delta \omega = T \eta_d S /\hbar \omega_p$. The output power $P_\mathrm{out}$ is constant at the optimal detuning $\Delta_{\mathrm{opt}}$ with
\begin{equation}
	P_\mathrm{out} = P_p\frac{\Gamma^2 - 3\Gamma\kappa + 3\kappa^2}{\Gamma^2}.
\end{equation} 
This leads to the following result of the minimum detectable frequency change or rather the LOD of
\begin{equation}
	\Delta \omega = \sqrt{\frac{\Gamma^2 - 3\Gamma\kappa + 3\kappa^2}{\Gamma^2} \cdot \frac{\hbar\omega}{\eta_d T P_p}} \frac{1}{\left| G_\mathrm{geo} G_\mathrm{det} \right|}\left(1-\frac{P_p}{P_\mathrm{max}}\right). \label{equ:minDetFreqRes}
\end{equation}
We display the calculated LOD in Fig. \ref{fig:LOD} assuming a linear and a nonlinear ring resonator with $g_\mathrm{tot}=0$ Hz and $g_\mathrm{tot}=112$ Hz, respectively. It can be seen that the LOD can actually be improved by exploiting the nonlinear SPM effect for a sensor application where the detector noise is the dominant noise source. The improvement then scales as $G^{-1}$ which is analogous to the scaling of the phase sensitivity.

\bibliography{SpmSensing_bib}

\end{document}